\documentclass[aps,tighten,amssymb,epsfig,showpacs]{revtex4}
\usepackage{epsfig}

\usepackage{amssymb}

\newcommand{\Real}{\mathbb{R}}

\begin{document}

\title{Critical bubbles and implications for critical black strings}

\author{Olivier Sarbach}
\affiliation{Department of Mathematics and
Department of Physics and Astronomy, Louisiana State University,
202 Nicholson Hall, Baton Rouge, Louisiana 70803-4001, USA}

\author{Luis Lehner}
\affiliation{Department of Physics and Astronomy, Louisiana State University,
202 Nicholson Hall, Baton Rouge, Louisiana 70803-4001, USA}

\begin{abstract}
We demonstrate the existence of gravitational critical phenomena in
higher dimensional electrovac bubble spacetimes. To this end, we study
linear fluctuations about families of static, homogeneous spherically
symmetric bubble spacetimes in Kaluza-Klein theories coupled to a
Maxwell field. We prove that these solutions are linearly unstable and
posses a unique unstable mode. We further show that the associated growth
rate  depends only on the mass per unit length of the background solution.
Additionally, by a double analytical continuation this mode can be seen
to correspond to marginally stable stationary modes of perturbed black
strings whose periods are integer multiples of the Gregory-Laflamme
critical length. This allow us to rederive recent results about the
behavior of the critical mass for large dimensions and to generalize
them to charged black string cases.
\end{abstract}

%


\maketitle

\section{Introduction}

Bubble spacetimes were originally studied for, among other reasons,
their connection with negative energy
solutions \cite{witten,brillpfister,brillhorowitz}. The behavior of
bubble spacetimes has been the subject of new scrutiny and use in
recent years. Among these, their use in defining new solutions of
Kaluza-Klein black
holes \cite{emparanreall,aharony,elvanghorowitz,elvangharmarkobers},
studies of possible phases of Kaluza-Klein black holes 
\cite{elvangharmarkobers}, solutions in inflationary universe in
five-dimensional spacetimes \cite{shiromizu} and their dynamical
behavior under different interactions via numerical simulations
\cite{sarbachlehner,sarbachlehner2}.

In particular, the numerical simulations presented in Ref.
\cite{sarbachlehner,sarbachlehner2} reveal strong indications of the
possible existence of critical phenomena in bubble spacetimes. That
is, there exists a critical value $k=k_c$ in the parameter specifying
the strength of the ${\bf U(1)}$ gauge field in the initial data for
the bubble. For this value, it was observed that, after some transient
behavior, the solution approaches a static bubble configuration which
we call the {\em critical bubble solution}. For values of $k$ close to $k_c$ 
the solution approaches the critical bubble and stays
close to it for some time and then either disperses away or collapses
to a black string. The amount of proper time $\tau$ that the solution
remains close to the static one obeys the relation $\tau/R_0 \simeq -
\Gamma\ln(|k-k_c|)$, where $R_0$ is the bubble location  --which is 
fixed by the ADM mass per unit length-- and $\Gamma$ is
a universal parameter that does not depend on the initial data.

In the present work we put on firm grounds such observation by
studying linear perturbations off the critical solutions in a $q+3$
dimensional spacetime endowed with ${\bf U(1)\times SO(q+1)}$ symmetry
and elucidating the existence of a unique unstable exponential in time
growing mode. Furthermore, we show that the growth rate $\Omega/R_0$
associated to this instability has the property that $\Omega$ does not
depend on the family of critical solutions and hence explains the
universal behavior. We also study the limit of large $q$'s and find
that in this case the exponent $\Omega$ obeys a simple power law rule.

Finally, our results can be used to draw conclusions on the `dual
black string system' obtained by appropriate double analytical
continuations. This maps the bubble solution into a charged black
string solution. Thus, exponential growing modes off the bubble
solution turn out to correspond to static, harmonic 'deformations' of
the uniform black strings. For these modes to exist,
the strings length must be long enough to accommodate it, and we find
that it must be an integer multiple of the critical length $L_c = 2\pi
R_0/\Omega$. $L_c$ results exactly that expected from the Gregory
Laflamme instability for the uncharged black strings\cite{GL}. For the
charged case, however, the black string solutions we found are not the 
same as those presented in Ref. \cite{GL2} as we here only consider ${\bf U(1)}$
gauge fields in the absence of a dilaton. 
Additionally, our results can be used to derive in a new
way the exponential law for the critical mass presented by Sorkin
\cite{sorkin} and to generalize it to the charged case. In particular,
in the latter case we show how the addition of charge worsens the situation
as far as stability of the string is concerned.\\

This work is organized as follows. In section \ref{Sect:Results} we
summarize the obtained results and discuss the implications for black
string spacetimes. The rest of the article presents the details of our
calculations. In section \ref{Sect:KKReduction} we consider
Kaluza-Klein theories coupled to a Maxwell field in spacetime
dimensions $q+3 \geq 5$ with ${\bf U(1) \times SO(q+1)}$ symmetry. We
perform a dimensional reduction of the action, obtaining an effective
action in two spacetimes dimension, and write down the equations of
motions. In section \ref{Sect:Static} we discuss the class of static
solutions which represent the critical bubbles. The linear stability
analysis about the bubble solutions follows in section
\ref{Sect:LinAnal} where we cast the perturbation equations into the
form of a wave equation whose spatial operator is self-adjoint. We
then show analytically that all bubble solutions are linearly
unstable, and give analytic arguments for the existence of a unique
non-degenerated eigenvalue $\lambda = -\Omega^2/R_0^2$ which is
negative. Furthermore, we show that for fixed spacetime dimension, the
corresponding frequency $\Omega$ is independent of the background
solution. We provide analytical estimates for $\Omega$ and compute it
by numerical means.

\section{Critical bubbles and black strings. Results and implications}
\label{Sect:Results}

\subsection{Critical bubbles: Summary of results}

In this article, we study the linear stability of the following
static bubble solutions which are solution of the Einstein-Maxwell
equations in $(q+3)$ spacetime dimensions:
\begin{eqnarray}
ds^2 &=& V^{2/q} \left[ -dt^2 + \frac{dR^2}{U} + R^2 \hat{g} \right] 
  + \frac{U}{V^2}\, dz^2,
\label{Eq:BubbleSol1}\\
A_\mu dx^\mu &=& P\frac{1-U}{V}\, dz,
\label{Eq:BubbleSol2}
\end{eqnarray}
where $P^2 = \epsilon(1-\epsilon)(q+1)/(q^2\kappa)$, $\kappa$ denoting
a coupling constant, and where
\begin{displaymath}
U = 1 - (R_0/R)^{q-1}, \qquad
V = 1 - \epsilon(R_0/R)^{q-1}.
\end{displaymath}
Here, $\hat{g}$ denotes the standard metric on the $q$-sphere $S^q$,
and the coordinate $R$ lies in the interval $[R_0,\infty)$. The
parameters $R_0 > 0$ (corresponding to the position of the bubble) and
$\epsilon\in [0,1)$ are related to the ADM mass and to the asymptotic
period $T$ of the extra dimension via $M_{ADM}=R_0^{q-1} T
|S^q|/16\pi$ and $T = 4\pi R_0(1-\epsilon)^{1 + 1/q}/(q-1)$.

\begin{figure}
\begin{center}
\epsfig{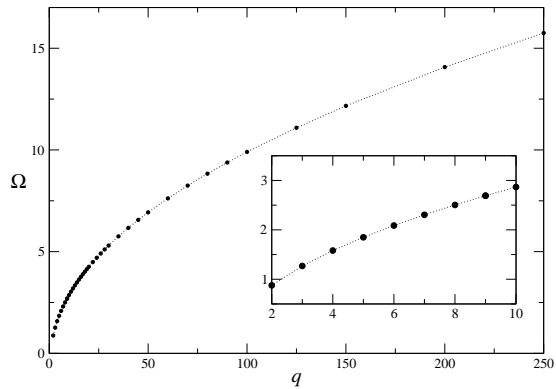}
\end{center}
\caption{The plot illustrates the obtained dependence of $\Omega$ on
$q$ (dark circles, with a dotted joining line for guidance). The inset 
shows $\Omega$ versus $q$ for low values of $q$.\\
\\}
\label{Fig:lam_q}
\end{figure}

\begin{figure}
\begin{center}
\epsfig{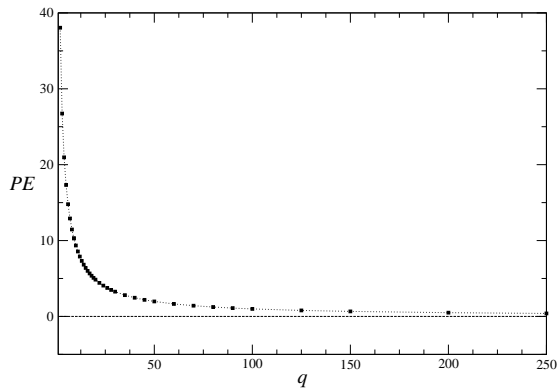}
\end{center}
\caption{Percentile error $PE$ of the actual value of $\Omega$ compared
to the simple law $\sqrt{q}$}
\label{Fig:error_q}
\end{figure}

We find spherically symmetric, in $z$ homogeneous, linear modes that
depart like $\exp(\Omega t/R_0)$ from the static solution. For the
uncharged case with $q=2$ this result can be drawn from the work
\cite{gross} though here we concern ourselves to more generic
cases. For each value of the spacetime dimension parameter $q$ we show
that there is a unique such mode and that the corresponding
dimensionless growth rate $\Omega$ does not depend on the parameter
$\epsilon$ which is determining the strength of the gauge field. This
{\em universal behavior} is precisely what is expected on physical
grounds for a critical solution\footnote{For a review on critical
phenomena in G.R.  see \cite{gundlachreview}}. It explains the
critical phenomena found in our numerical simulations
\cite{sarbachlehner,sarbachlehner2} where the universal parameter
$\Gamma$ is related to $\Omega$ via $\Gamma = 1/\Omega$, i.e. $\Gamma$
is the inverse of the Lyapunov exponent of the single unstable
mode. Table \ref{Tab:Omega} and figure \ref{Fig:lam_q} show the
obtained numerical results for $\Omega$ as a function of $q$.
Furthermore, our analysis reveals that that for $q$ large enough,
$\Omega$ is approximately given by $\Omega \simeq q^{1/2}$. Figure
\ref{Fig:error_q} displays the percentile error $PE= 100 \, |\Omega -
q^{1/2}|/q^{1/2}$.

\begin{table}[h]
\center
\begin{tabular}{|l||c|c|c|c|c|c|c|c|c|c|c|c|c|c|}\hline
 $q$      & $2$ & $3$ & $4$ & $5$ & $6$ & $7$ & $8$ & $9$ & $10$ & $20$ & $30$ & $40$ & $50$ & $100$\\ 
\hline
$\Omega$  & $0.876$ & $1.27$ & $1.58$ & $1.85$ & $2.09$ & $2.30$ & $2.50$ & $2.69$ & $2.87$ & $4.26$ & $5.30$ & $6.17$ & $6.93$ & $9.90$ \\ 
\hline
\end{tabular}
\caption{Numerical values for $\Omega$ for different values of $q$.}
\label{Tab:Omega}
\end{table}

Summarizing, {\em bubble spacetimes} admit the critical solutions
(\ref{Eq:BubbleSol1},\ref{Eq:BubbleSol2}) for spacetime dimensions
$q+3 \geq 5$ with corresponding critical exponent $\Omega$ which is
independent of $\epsilon$. Furthermore, for $q$ large enough, $\Omega
\simeq q^{1/2}$. Indeed, even at $q = 7$ this approximation differs
only by $\simeq 15$\% from the actual critical value.

\subsection{Implications for black strings}

Connections between possible black string instabilities and modes in a
`dual' space obtained by Wick rotations have been noted, and employed,
in the past (see Refs. \cite{gross,GL0,reall}). We here illustrate how
one can make the connection between bubble and black string spacetimes
via double-Wick rotations to draw conclusions for the latter case.

The bubble configurations (\ref{Eq:BubbleSol1},\ref{Eq:BubbleSol2})
are related to a family of charged black string solutions via the
double analytical continuation $t \mapsto i z$, $z \mapsto i t$
followed by the transformation $\epsilon \mapsto -\epsilon$:
\begin{eqnarray}
ds^2 &=& -\frac{U}{V^2} dt^2 + V^{2/q}\left( \frac{dR^2}{U} + dz^2 + R^2\hat{g} \right),
\label{Eq:StatMetricBS}\\
A_\mu dx^\mu &=& -Q\frac{1-U}{V}\, dt,
\label{Eq:StatGFBS}
\end{eqnarray}
where now $U = 1 - (R_0/R)^{q-1}$, $V = 1 + \epsilon(R_0/R)^{q-1}$,
and $Q^2 = \epsilon(1+\epsilon)(q+1)/(q^2\kappa)$. The parameters
$R_0$ (which now corresponds to the location of the horizon) and
$\epsilon$ satisfy $R_0 > 0$, $\epsilon \geq 0$ and are related to the
mass and electric charge via 
\begin{eqnarray}
M &=& \frac{R_0^{q-1} L |S^q|}{16\pi}
\left[ q + 2\epsilon\left(q - \frac{1}{q} \right) \right], \\
Q_e &=& \frac{(q-1)Q R_0^{q-1} L |S^q|}{4\pi}\; ,
\nonumber
\end{eqnarray}
where $L$ denotes the asymptotic period of the extra dimension. For
$q=2$ these solutions agree with the solutions described by
Eqs. (2.13) and (2.15) of Ref. \cite{horowitzmaeda_solution}.

Under the double analytical continuation, the unstable mode of the
critical bubbles which is proportional to $e^{\Omega t/R_0}$ is mapped
to a stationary mode of the corresponding black strings which is
proportional to $e^{i\Omega z/R_0}$. This mode represents a static
deformation of the black string, with harmonic dependency in
$z$. That the existence of static modes are central in the understanding
of instabilities has been discussed in \cite{gross,reall}, we here exploit
knowing these modes's wavelengths explicitly to draw conclusions of
 black string instabilities. 
Clearly, for this mode to exist, the asymptotic periodicity of
the extra dimension must be an integer multiple of $L_c = 2\pi
R_0/\Omega$. In the uncharged case, it turns out that $L_c$ is exactly
equal to the critical length for the Gregory-Laflamme instability
\cite{GL,GL2}. (Compare the values of Table \ref{Tab:Omega} with the
values in Table 1 of Ref. \cite{sorkinkol}.) 
For the charged case the critical solution does not correspond to the
charged black string background solutions considered in\cite{GL2} as
ours only include the interaction with a $U(1)$ gauge field. However,
as in~\cite{GL2} we expect that the presence of a static deformation 
is related to a threshold between stable and unstable solutions.

We are able to give the dependency for the critical length in terms of the
parameter $\epsilon$ in an analytic way since we have shown that
$\Omega$ is independent on $\epsilon$. The result is conveniently
expressed in terms of the dimensionless mass defined in \cite{sorkin}
\begin{equation}
\mu := \frac{M_c}{L_c^q} = \frac{|S^q|}{16\pi} \left( \frac{\Omega}{2\pi} \right)^{q-1}
\left[ q + 2\epsilon\left(q - \frac{1}{q} \right) \right].
\end{equation}
For large $q$ we can approximate $\Omega \simeq q^{1/2}$ and find
\begin{equation}
\mu \simeq \frac{\sqrt{q}}{16} \left( \frac{e}{2\pi} \right)^{q/2} (1 + 2\epsilon),
\end{equation}
which provides a different way of deriving the law presented in
Ref. \cite{sorkin,sorkinkol}, and generalizes it to the 
electrically charged case.

Note that in the charged case, the string becomes {\it ``more unstable''} in
the sense that the critical length at which instabilities arise is shorter
that the charge-free case. This is to be contrasted with the case studied
in \cite{GL2,GL3} where the opposite occurs. In that case, however, the
interaction with a magnetic field was considered while in the present an
electric field is included instead.

\section{Kaluza-Klein reduction and effective action}
\label{Sect:KKReduction}

We consider a $q+3$ dimensional spacetime $(M,g)$ with ${\bf U(1)
\times SO(q+1)}$ symmetry. We assume that the Killing field
$\partial_z$ which is the generator of the action of the ${\bf U(1)}$
group on $M$ is hypersurface orthogonal. The manifold has the
structure $M = \tilde{M} \times S^q \times S^1$ with metric
\begin{equation}
ds^2 = e^{-2\phi/q}\left[ \tilde{g}_{ab} dx^a dx^b + r^2\hat{g} \right] + e^{2\phi} dz^2,
\label{Eq:KKMetric}
\end{equation}
where $\phi$, $r$ and $\tilde{g}_{ab} dx^a dx^b$ are, respectively, a
function, a positive function, and a pseudo-Riemannian metric on the
two-dimensional manifold $\tilde{M}$. $\hat{g}$ denotes the standard
metric on the $q$-dimensional sphere $S^q$. In this article, we assume
that $q \geq 2$.  We also consider a ${\bf U(1)}$ gauge potential of
the form
\begin{equation}
A_\mu dx^\mu = \gamma\, dz,
\end{equation}
with $\gamma$ a function on $\tilde{M}$.

The dynamics is described by the $q+3$ dimensional
Einstein-Hilbert-Maxwell action
\begin{equation}
S = -\frac{1}{16\pi}\int \left[ Ric - q\kappa\, g(dA,dA) \right] \sqrt{-g}\, d^{q+3} x,
\end{equation}
where $Ric$ denotes the Ricci scalar belonging to the metric
(\ref{Eq:KKMetric}) and $\kappa$ is a coupling constant. When
integrating over $z$, one obtains a reduced action describing $q+2$
dimensional gravity coupled to the dilaton field $\phi$ and to the
${\bf U(1)}$ gauge potential $\gamma\, dz$.  Using the $SO(q+1)$
symmetry one can further reduce the action\footnote{see, for instance,
Ref. \cite{BHS} for a derivation in the case $q=2$.} and, after some
manipulations, obtains
\begin{equation}
S_{eff} = -\int \left[ 2\frac{\tilde{g}(dm,dr)}{N} - \frac{q+1}{q^2}\,r^q\tilde{g}(d\phi,d\phi)
 - \kappa\; r^q e^{-2\phi}\tilde{g}(d\gamma,d\gamma) \right] \sqrt{-\tilde{g}}\, dx^0\wedge dx^1,
\end{equation}
where the functions $N$ and $m$ are defined by $N = \tilde{g}(dr,dr)$,
$N = 1 - 2m\, r^{1-q}$, and where $(x^0,x^1)$ are local coordinates on
$\tilde{M}$. We stress that up to this point all the fields are
defined in an invariant geometrical way: $e^{2\phi}$ is the norm of
the Killing field $\partial_z$, $r$ is defined via the area radius,
$e^{-\phi/q} r$, of the $q$-spheres, and $N$ is the norm of $dr$.

In order to write down the field equations, we choose the coordinates
$(x^0,x^1)$ such that $x^1=r$, set $t=x^0$ and parametrize the
two-metric according to
\begin{equation}
\tilde{g}_{ab} dx^a dx^b = -\sigma^2 N (dt + a\, dr)^2 + \frac{dr^2}{N}\; ,
\end{equation}
where $\sigma = \sqrt{-\tilde{g}}$ and $a = -\tilde{g}(dt,dr)/N$.
The action has the form
\begin{equation}
S_{eff} = -2\int \left[ {\cal L}_G + {\cal L}_M \right] dt \wedge dr,
\label{Eq:EffAction}
\end{equation}
with
\begin{eqnarray}
{\cal L}_G &=& \sigma(m' - a\dot{m}), \\
{\cal L}_M &=& \frac{\sigma r^q}{2}\left[ \frac{\dot{\psi}^2}{\sigma^2 N} - N(\psi' - a\dot{\psi})^2 \right]
 + \frac{\kappa\,\sigma r^q e^{-\alpha\psi}}{2} \left[ \frac{\dot{\gamma}^2}{\sigma^2 N} - N(\gamma' - a\dot{\gamma})^2 \right],
\end{eqnarray}
where we have defined $\psi = \sqrt{(q+1)/q^2}\, \phi$, $\alpha =
2q/\sqrt{q+1}$ for convenience, and where a dot and a prime denote
differentiation with respect to $t$ and $r$, respectively.

Of course, we can always choose the coordinate $t$ such that
$a=0$. However, we shall set $a=0$ only {\em after} varying the action
since otherwise an equation is lost. The field equations obtained from
varying (\ref{Eq:EffAction}) with respect to $\sigma$, $m$, $a$,
$\psi$ and $\gamma$ are, after setting $a=0$,
\begin{eqnarray}
2m' &=& r^q\left[ \frac{\dot{\psi}^2}{\sigma^2 N} + N\psi'^2 \right]
 + \kappa\, r^q e^{-\alpha\psi} \left[ \frac{\dot{\gamma}^2}{\sigma^2 N} + N\gamma'^2 \right],
\label{Eq:mprime}\\
N\frac{\sigma'}{\sigma} &=&  r\left[ \frac{\dot{\psi}^2}{\sigma^2 N} + N\psi'^2 \right]
 + \kappa\, r e^{-\alpha\psi} \left[ \frac{\dot{\gamma}^2}{\sigma^2 N} + N\gamma'^2 \right],
\label{Eq:sigma}\\
\dot{m} &=& r^q N\dot{\psi}\psi' + \kappa r^q N e^{-\alpha\psi}\dot{\gamma}\gamma',
\label{Eq:mdot}
\end{eqnarray}
and
\begin{eqnarray}
&& r^q\partial_t\left( \frac{\dot{\psi}}{\sigma N} \right) - \left( r^q \sigma N \psi' \right)' 
 = -\frac{\alpha\kappa}{2}\, \sigma r^q e^{-\alpha\psi}\left[ \frac{\dot{\gamma}^2}{\sigma^2 N} - N\gamma'^2 \right],
\label{Eq:psi}\\
&& r^q\partial_t\left( e^{-\alpha\psi} \frac{\dot{\gamma}}{\sigma N} \right) - \left( r^q \sigma N e^{-\alpha\psi} \gamma' \right)' = 0.
\label{Eq:gamma}
\end{eqnarray}
Combining Eqs. (\ref{Eq:mprime}) and (\ref{Eq:sigma}) we also find
\begin{equation}
\frac{1}{\sigma} (\sigma N)' = (q-1)\frac{2m}{r^q}\, .
\label{Eq:Deltar}
\end{equation}
In the next section, we discuss static solutions to these equations,
where $\dot{m} = \dot{\psi} = \dot{\gamma} = 0$.

\section{Static solutions}
\label{Sect:Static}

When $\gamma = \psi = 0$ we obtain $m = const$, so the only solutions
to the field equations consists of the one-parameter family of $q+2$
dimensional Schwarzschild spacetimes cross $S^1$. For static
configurations in general, we notice that Eqs. (\ref{Eq:psi}),
(\ref{Eq:gamma}) posses the first integrals
\begin{eqnarray}
&& r^q\sigma N e^{-\alpha\psi} \gamma' = \tilde{P},
\label{Eq:Int1}\\
&& r^q\sigma N\psi' + \frac{1}{2}\,\alpha\kappa\tilde{P} \gamma = \kappa\tilde{Q},
\label{Eq:Int2}
\end{eqnarray}
with two constants $\tilde{P}$, $\tilde{Q}$. In fact, the static
equations can be integrated completely for bubble configurations
obtaining the following two-parameter family of solutions,
\begin{eqnarray}
ds^2 &=& \frac{V^{2/q}}{U^{1/q}}\left[ -U^{1/q} dt^2 + U^{1/q-1} dR^2 + R^2 U^{1/q}\hat{g} \right] + \frac{U}{V^2}\, dz^2,
\label{Eq:StatMetric}\\
A_\mu dx^\mu &=& \frac{P}{\epsilon V}\, dz,
\label{Eq:StatGF}
\end{eqnarray}
where $P^2 = \epsilon(1-\epsilon)(q+1)/(q^2\kappa)$, and where
\begin{displaymath}
U = 1 - (R_0/R)^{q-1}, \qquad
V = 1 - \epsilon(R_0/R)^{q-1}.
\end{displaymath}
The coordinate $R$ lies in the open interval $[R_0,\infty)$.  The
parameters $R_0 > 0$ (corresponding to the position of the bubble) and
$\epsilon\in [0,1)$ are related to the ADM mass and to the asymptotic
period $T$ of the extra dimension, see section \ref{Sect:Results}.  In
order to determine the latter, we replace the coordinate $R$ by the
coordinate $\hat{\rho}$ defined by $U = \hat{\rho}^2$ and rewrite
\begin{displaymath}
\frac{V^{2/q}}{U}\, dR^2 + \frac{U}{V^2}\, dz^2
 = \frac{1}{V^2}\left( \frac{4 R_0^2}{(q-1)^2} (1-\epsilon+\epsilon\hat{\rho}^2)^{\frac{2(q+1)}{q}}
 (1-\hat{\rho}^2)^{-\frac{2q}{q-1}} d\hat{\rho}^2 
 + \hat{\rho}^2 dz^2 \right).
\end{displaymath}
Therefore, in order to avoid a conical singularity at $\hat{\rho}=0$,
we need $T = 4\pi R_0(1-\epsilon)^{1 + 1/q}/(q-1)$. The integration
constants $\tilde{P}$, $\tilde{Q}$ are related to $R_0$ and $P$ via
\begin{displaymath}
\tilde{P} = -(q-1)R_0^{q-1} P, \qquad
\tilde{Q} = -\frac{(q-1)R_0^{q-1}}{\alpha\kappa}\; .
\end{displaymath}

In the next section we consider time-dependent bubble configurations
that are infinitesimally close to the static solutions. That is, we
consider metrics of the form
\begin{displaymath}
ds^2 = \left( \frac{V^2}{U} e^{-2E} \right)^{1/q}
\left[ -U^{1/q} e^{2A} dt^2 + U^{1/q-1} e^{2B} dR^2 + R^2\, U^{1/q} e^{2C}\hat{g} \right] + \frac{U}{V^2} e^{2E} dz^2,
\end{displaymath}
where $A$, $B$, $C$, $E$ are ``small'' functions of $t$ and $R$, which
vanish as $R \rightarrow\infty$ and such that $B(t,R_0)= (1 +
1/q)E(t,R_0)$.  The former conditions are necessary for asymptotic
flatness while the latter ensures that the period of the extra
dimension $T$ does not change. In order to determine the asymptotic
behavior in the amplitudes $\sigma$, $N$, $\psi$ and $\gamma$ we
notice that
\begin{eqnarray}
e^{\alpha\psi} &=& \frac{U}{V^2} e^{2E},\\
r &=& R\, U^{1/2q} e^C,\\
N &=& \tilde{g}(dr,dr) = U^{1-1/q} e^{-2B} \left[ \frac{\partial r}{\partial R} \right]^2,\\
\sigma^2 N &=& U^{1/q} e^{2A},\\
\gamma &=& \frac{P}{\epsilon V} + \delta\gamma.
\end{eqnarray}
Clearly, this implies that $\psi\rightarrow 0$, $\sigma, N \rightarrow
1$ as $R \rightarrow\infty$. In order to analyze the behavior near the
bubble, we introduce the dimensionless quantity $x = (R - R_0)/R_0$.
The background quantities (the ones obtained by setting $A=B=C=E=0$)
have the expansion
\begin{eqnarray}
e^{\alpha\psi} &=& \frac{(q-1)x}{(1-\epsilon)^2} + O(x^2),\\
r^{2q} &=& R_0^{2q} (q-1) x + O(x^2),\\
\sigma &=& 2q (q-1)^{-\frac{q-1}{2q}} x^{\frac{q+1}{2q}} \left[ 1 + O(x) \right], \\
N &=& \frac{(q-1)}{4q^2} \frac{1}{x} + O(1).
\end{eqnarray}
Notice that $N$ diverges as $R \rightarrow R_0$, but $\sigma^2 N$ goes to zero.
Linearizing about the background, we find that $\delta C = 0$ (since we do not
vary $r$), and $\delta\psi = 2\alpha^{-1}\delta E$, $\delta N = -2N\delta B$.
As a consequence of this,
\begin{equation}
\left. \sigma\delta m \right|_{R=R_0} = \frac{q-1}{2q} R_0^{q-1} \delta B(t,R_0) 
 = \frac{q^2-1}{2q^2} R_0^{q-1} \delta E(t,R_0).
\label{Eq:sigmadmbub}
\end{equation}
Here and in the following, $\delta(...)$ denotes the linearization of
the quantity $(...)$ about a static background.

\section{Linear perturbations}
\label{Sect:LinAnal}

Since when $\gamma = \psi = 0$ all solutions are static, and in view
of the wave-like character of the equations (\ref{Eq:psi}),
(\ref{Eq:gamma}) for $\psi$ and $\gamma$, we expect that the true
dynamical degrees of freedom are contained in $\gamma$ and $\psi$. It
turns out that this expectation is indeed true at least at the
perturbative level: Our effective action Eq. (\ref{Eq:EffAction})
satisfies the form of Ref. \cite{BHS} where it was shown that for
linear perturbations around a static background the ``gravitational''
degrees of freedom can be completely eliminated and a pulsation
equation for the perturbed matter fields alone can be obtained. In
order to see this explicitly, we first linearize Eqs. (\ref{Eq:mdot})
and (\ref{Eq:mprime}), obtaining
\begin{eqnarray}
\delta\dot{m} &=& r^q N\left( \psi'\delta\dot{\psi} + \kappa e^{-\alpha\psi} \gamma'\delta\dot{\gamma} \right),
\label{Eq:mdotlin}\\
\delta m' &=& r^q N\left( \psi'\delta\psi' + \kappa e^{-\alpha\psi} \gamma'\delta\gamma' \right) 
 + \frac{r^q\delta N}{2} \left( \psi'^2 + \kappa e^{-\alpha\psi}\gamma'^2 \right)
 - \frac{\alpha\kappa}{2}\, r^q N e^{-\alpha\psi} \gamma'^2\delta\psi.
\end{eqnarray}
Using $r^q \delta N = -2r\delta m$ and the background equation
$\sigma'/\sigma = r\psi'^2 + \kappa r e^{-\alpha\psi}\gamma'^2$ we can rewrite the second equation as
\begin{equation}
[ \sigma\delta m]' = r^q\sigma N\left( \psi'\delta\psi' + \kappa e^{-\alpha\psi} \gamma'\delta\gamma' \right) 
 - \frac{\alpha\kappa}{2}\, r^q \sigma N e^{-\alpha\psi}\gamma'^2\delta\psi.
\end{equation}
Finally, using the static version of Eqs. (\ref{Eq:psi}),
(\ref{Eq:gamma}), we find
\begin{equation}
[ \sigma\delta m]' = \left[ r^q\sigma N\left( \psi'\delta\psi 
 + \kappa e^{-\alpha\psi} \gamma'\delta\gamma \right) \right]'.
\label{Eq:mprimlin}
\end{equation}
Equations (\ref{Eq:mdotlin}) and (\ref{Eq:mprimlin}) imply that
\begin{equation}
\sigma\delta m = r^q\sigma N\left( \psi'\delta\psi 
 + \kappa e^{-\alpha\psi} \gamma'\delta\gamma \right) + const.\, ,
\end{equation}
which, using Eqs. (\ref{Eq:Int1}) and (\ref{Eq:Int2}), can also be written as
\begin{equation}
\sigma\delta m = \kappa\left( \tilde{Q} - \frac{1}{2}\, \alpha\tilde{P}\gamma \right)\delta\psi + \tilde{P}\delta\gamma + const.
\label{Eq:sigmadm}
\end{equation}
In order to determine the constant, we evaluate this equation at
$R=R_0$.  Taking into account Eq. (\ref{Eq:sigmadmbub}) we obtain $0 =
\tilde{P}\delta\gamma(t,R_0) + const$. Since the addition of a
constant to $\delta\gamma$ is a gauge transformation, we set the
constant to zero in the following, which is equivalent to requiring
$\delta\gamma(t,R_0)=0$. Using Eq. (\ref{Eq:sigmadm}) and the
linearization of Eq. (\ref{Eq:Deltar}), we also find
\begin{equation}
\delta\left[ \frac{ (\sigma N)' }{\sigma N} \right] = \frac{2(q-1)}{N}\left( \psi'\delta\psi 
 + \kappa e^{-\alpha\psi} \gamma'\delta\gamma \right).
\label{Eq:LinGrav}
\end{equation}
Eq. (\ref{Eq:LinGrav}) allow to reexpress all gravitational
perturbations in terms of matter perturbations in the linearized
versions of Eqs. (\ref{Eq:psi}) and (\ref{Eq:gamma}).  Therefore, as
anticipated, the linearized ``gravitational'' degrees of freedom can
be entirely expressed in terms of the linearized ``matter'' fields,
and one obtains the following pulsation equation
\begin{equation}
{\bf P} \ddot{v} - ({\bf N}^T)^{-1}\sigma N\partial_r 
 \left[ \sigma N {\bf N}^T {\bf P} {\bf N}\partial_r \left( {\bf N}^{-1} v \right) \right] + {\bf S} v = 0,
\label{Eq:PulsEq}
\end{equation}
with
\begin{eqnarray}
{\bf P} &=& r^q\left( \begin{array}{cc} 1 & 0 \\ 0 & e^{-\alpha\psi} \end{array} \right),
\nonumber\\
{\bf S} &=& -\frac{\alpha^2\kappa\tilde{P}^2 e^{\alpha\psi}}{2 r^q} \left( \begin{array}{cc} 1 & 0 \\ 0 & 0 \end{array} \right)
 - \frac{2(q-1)}{r^q N} \left( \begin{array}{c} \kappa\tilde{Q} - \frac{1}{2}\,\alpha\kappa\tilde{P}\gamma \\ \sqrt{\kappa}\tilde{P} \end{array} \right)
 \left( \kappa\tilde{Q} - \frac{1}{2}\,\alpha\kappa\tilde{P}\gamma , \sqrt{\kappa}\tilde{P} \right),
\nonumber\\
{\bf N} &=& \left( \begin{array}{cc} 1 & 0 \\ \alpha\sqrt{\kappa}\gamma & 1 \end{array} \right),
\nonumber
\end{eqnarray}
and where $v = (\delta\psi, \sqrt{\kappa}\delta\gamma)^T$. The spatial
part of the operator acting on $v$ in Eq. (\ref{Eq:PulsEq}) is a
formally self-adjoint elliptic operator. In the following, we start
with a qualitative analysis of this operator, introducing a Hilbert
space on which it is defined, and giving estimates for the lower limit
of its spectrum. Then, we prove the existence of a unique bound state,
i.e. an eigenfunction belonging to a negative eigenvalue. In this
sense, each bubble configuration possesses precisely one unstable
mode. The timescale associated to this mode explains the critical
exponent in the critical dynamical behavior that has been announced in
Ref. \cite{sarbachlehner} and that will be discussed in detail in
Ref. \cite{sarbachlehner2}.

\subsection{Qualitative analysis of the pulsation operator}

Here we give a first analysis of the pulsation operator which is
defined by the spatial part of the operator acting on $v$ in
Eq. (\ref{Eq:PulsEq}). In order to do so, we first rewrite the
pulsation equations in terms of the dimensionless quantities $\bar{R}
= R/R_0$, $\bar{t} = t/R_0$, $q$ and $\epsilon$. Using the relations
in section \ref{Sect:Static}, we find
\begin{equation}
\left( \frac{\partial^2}{\partial\bar{t}^2} + {\cal A} \right) v = 0,
\label{Eq:PulsEqDimLess}
\end{equation}
where the pulsation operator is defined by
\begin{equation}
{\cal A} = -{\bf \bar{P}}^{-1} ({\bf N}^T)^{-1}\frac{\partial}{\partial\bar{R}} 
 \left[ U {\bf N}^T {\bf \bar{P}} {\bf N}\frac{\partial}{\partial\bar{R}} \left( {\bf N}^{-1} . \right) \right] + {\bf \bar{P}}^{-1} {\bf \bar{S}}\, ,
\end{equation}
where
\begin{eqnarray}
{\bf \bar{P}} &=& \bar{R}^q\left( \begin{array}{cc} 1 & 0 \\ 0 & \frac{V^2}{U} \end{array} \right),
\nonumber\\
{\bf \bar{S}} &=& -\frac{2(q-1)^2}{\bar{R}^q}\frac{\epsilon(1-\epsilon)}{V^2} \left( \begin{array}{cc} 1 & 0 \\ 0 & 0 \end{array} \right)
 - \frac{2(q^2-1)}{\bar{R}^q\left[ 1 + \frac{q+1}{q-1} U \right]^2} 
 \left( \begin{array}{c} 1 - 2(1-\epsilon)V^{-1} \\ 2\sqrt{\epsilon(1-\epsilon)} \end{array} \right)
 \left(  1 - 2(1-\epsilon)V^{-1}, 2\sqrt{\epsilon(1-\epsilon)} \right),
\nonumber\\
{\bf N} &=& \left( \begin{array}{cc} 1 & 0 \\ 2\sqrt{\frac{1-\epsilon}{\epsilon}} V^{-1} & 1 \end{array} \right),
\nonumber
\end{eqnarray}
and where
\begin{displaymath}
U = 1 - \bar{R}^{-(q-1)}, \qquad
V = 1 - \epsilon\bar{R}^{-(q-1)}.
\end{displaymath}

In order to discuss the pulsation equation, we define an appropriate
Hilbert space in which the operator ${\cal A}$ (defined on a suitable
domain) is self-adjoint. The solutions to Eq. (\ref{Eq:PulsEqDimLess})
can then be constructed using spectral theory, and the long-time
behavior of the solutions is determined by the spectrum of ${\cal A}$.
We define the Hilbert space
\begin{displaymath}
{\cal H} = L^2( (1,\infty),\Real^2; {\bf \bar{P}} d\bar{R} )
\end{displaymath}
of all functions $v: (1,\infty) \rightarrow \Real^2$ which are square
integrable with respect to the measure ${\bf \bar{P}} d\bar{R}$. The
corresponding scalar product is given by
\begin{equation}
(u, v) = \int_1^\infty u^T {\bf \bar{P}} v\, d\bar{R}.
\label{Eq:ScalProd}
\end{equation}
Let $u,v$ be smooth functions on $(1,\infty)$. Using integration by
parts we obtain
\begin{equation}
(u, {\cal A}v ) = \int_1^\infty \left[ \frac{\partial}{\partial\bar{R}} ({\bf N}^{-1} u)^T \cdot U 
 {\bf N}^T {\bf \bar{P}} {\bf N}\frac{\partial}{\partial\bar{R}} \left( {\bf N}^{-1} v \right) 
+ u^T {\bf \bar{S}} v \right] d\bar{R}
- \left. u^T U{\bf \bar{P}} {\bf N} \frac{\partial}{\partial\bar{R}} \left( {\bf N}^{-1} v \right)  \right|_{\bar{R}=1}^\infty.
\label{Eq:uAv}
\end{equation}
Defining the domain $D({\cal A})$ of ${\cal A}$ to be the set of all
smooth functions $u: (1,\infty) \rightarrow \Real^2$ such that all
derivatives are uniformly bounded on $(1,\infty)$ and such that
$u_2(\bar{R}=1)=0$ and $u$ and its derivatives decay faster than
$\bar{R}^{-q/2-1/2}$ as $\bar{R}\rightarrow\infty$, we see that the
boundary term vanishes for $u,v\in D({\cal A})$, and so
\begin{equation}
(u, {\cal A}v ) = ({\cal A}u, v)
\end{equation}
for all $u,v\in D({\cal A})$. Therefore, ${\cal A}: D({\cal A})
\subset {\cal H} \rightarrow {\cal H}$ is a densely defined, symmetric
operator. Since ${\cal A}$ is real, in the sense that it commutes with
complex conjugation, ${\cal A}$ can be extended to a self-adjoint
operator; that is, ${\cal A}$ can be defined on a larger domain
$D(\hat{\cal A}) \supset D({\cal A})$ to yield a self-adjoint operator
$\hat{\cal A}: D(\hat{\cal A}) \subset {\cal H} \rightarrow {\cal H}$.
In general, there might exist different such self-adjoint
extensions. If ${\cal A}$ is bounded from below, one can single out an
extension by requiring that it possesses the same lower bound as
${\cal A}$. In the following, we show that ${\cal A}$ is bounded from
below, and take as extension $\hat{\cal A}$ this unique self-adjoint
extension. We refer the reader to Ref. \cite{ReedSimon} for the theory
of self-adjoint extensions of densely defined symmetric operators.

In order to show that ${\cal A}$ is bounded from below, and in order
to give an estimate for this lower bound, we estimate the quantity
\begin{equation}
G \equiv \inf\limits_{v\in D({\cal A})\atop v\neq 0} \frac{ (v,{\cal A}v) }{ (v,v) } \; .
\end{equation}
We first show that $G > -\infty$ which implies that ${\cal A}$ is
bounded from below and so there is a unique self-adjoint extension
$\hat{\cal A}$ which has the same lower bound $G$. We will also show
that $G$ is negative which means that the background solutions are
unstable. Furthermore, we will show that $\hat{\cal A}$ possesses
precisely one eigenfunction with negative eigenvalue. In this
subsection we provide estimates for this eigenvalue; its existence and
uniqueness and the numerical computation of its actual value are given
in the next subsection.

Our analysis of the pulsation operator ${\cal A}$ is greatly
simplified by performing the transformation $v \mapsto {\bf O} v$
where the matrix ${\bf O}$ is given by
\begin{equation}
{\bf O} = \left( \begin{array}{cc} 
 1 - 2(1-\epsilon)V^{-1} & -2\sqrt{\epsilon(1-\epsilon)} \\ 
 2\sqrt{\epsilon(1-\epsilon)}\, U V^{-2} & 1 - 2(1-\epsilon)V^{-1} \end{array} \right).
\end{equation}
Notice that ${\bf O}$ defines an orthogonal transformation with respect to
the scalar product defined in (\ref{Eq:ScalProd}) since ${\bf O}^T
{\bf\bar{P}} {\bf O} = {\bf{\bar P}}$. Furthermore, ${\bf O}$
leaves the domain $D({\cal A})$ of ${\cal A}$ invariant. Remarkably, the
transformation ${\bf O}$ brings the pulsation operator into diagonal form:
\begin{equation}
{\bf O}^\dagger {\cal A} {\bf O} = {\bf\bar{P}}^{-1} 
\left( \begin{array}{cc} {\cal B}_1 & 0 \\ 0 & {\cal B}_2 \end{array} \right),
\end{equation}
where ${\bf O}^\dagger = {{\bf{\bar P}}^{-1} {\bf O}^T \bf{\bar P}} = {\bf O}^{-1}$ and
\begin{eqnarray}
{\cal B}_1 &=& -\frac{\partial}{\partial\bar{R}} U \bar{R}^q \frac{\partial}{\partial\bar{R}} 
 - \frac{2(q^2-1)}{\bar{R}^q\left[ 1 + \frac{q+1}{q-1} U \right]^2}\; ,
\\
{\cal B}_2 &=&  -\frac{\partial}{\partial\bar{R}} V^2 \bar{R}^q \frac{\partial}{\partial\bar{R}}\; .
\end{eqnarray}
Therefore, it is sufficient to analyze the two decoupled eigenvalue
problems ${\cal B}_1 u_1 = \lambda \bar{R}^q u_1$, ${\cal B}_2 u_2 =
\lambda \bar{R}^q V^2 U^{-1} u_2$, $(u_1,u_2)^T \in D({\cal A})$, with
$\lambda\in\Real$ an eigenvalue. Since ${\cal B}_2$ is a positive
operator, only the operator ${\cal B}_1$ can lead to
instabilities. Notice that ${\cal B}_1$ does not depend on $\epsilon$.
{\em Therefore, the bound states of the pulsation operator (if they
exist) are independent of $\epsilon$.} In the following, we derive
lower and upper bounds for the quantity
\begin{equation}
G_1 \equiv \inf\limits_{u\in D({\cal A})\atop u\neq 0} \frac{ \int_1^\infty u_1 {\cal B}_1 u_1\, d\bar{R} }{ \int_1^\infty \bar{R}^q u_1^2\, d\bar{R} }\; .
\end{equation}
Since the transformation ${\bf O}$ is orthogonal, ${\cal B}_2$ is
positive, and since we will show that $G_1$ is negative, it follows
that $G = G_1$, so $G_1$ gives the lower limit of the spectrum of the
pulsation operator.

Using integration by parts, we find, for $u\in D({\cal A})$:
\begin{eqnarray}
\int_1^\infty u_1 {\cal B}_1 u_1\, d\bar{R} &=& \int_1^\infty \left[ U\bar{R}^q \left( \frac{\partial u_1}{\partial\bar{R}} \right)^2 
 - \frac{2(q^2-1) u_1^2}{\bar{R}^q\left[ 1 + \frac{q+1}{q-1} U \right]^2} \right] d\bar{R}
\nonumber\\
 &\geq& -2(q^2-1) \int_1^\infty \bar{R}^q u_1^2 d\bar{R},
\nonumber
\end{eqnarray}
which shows that $G_1 \geq -2(q^2-1)$. Next, we derive upper bounds
for $G_1$. In order to do so, we introduce the quantities
\begin{eqnarray}
Z &=& (q-1) \int_1^\infty \bar{R}^q u_1^2 d\bar{R} 
   = \int_0^1 (1-U)^{-\frac{2q}{q-1}} u_1^2 dU,
\nonumber\\
E &=& \frac{1}{q-1} \int_1^\infty u_1 {\cal B}_1 u_1\, d\bar{R} 
   = \int_0^1 \left[ U \left( \frac{\partial u_1}{\partial U} \right)^2 
   - 2\frac{q+1}{q-1}\frac{u_1^2}{\left[ 1 + \frac{q+1}{q-1} U \right]^2} \right] dU,
\nonumber 
\end{eqnarray}
and minimize the functional $u_1 \mapsto E/Z$ over suitable test
functions. We first study the limit $q\rightarrow\infty$. In this
limit, $E$ becomes positive. Indeed, using integration by parts and
requiring $Z < \infty$, we can rewrite
\begin{displaymath}
E = \int_0^1 U \left[ \frac{1-U}{1+U}\frac{\partial}{\partial U}
 \left( \frac{1+U}{1-U} u_1 \right) \right]^2 dU.
\end{displaymath}
Therefore, in the limit $q \rightarrow\infty$, $E \geq 0$ and $E=0$ if
and only if $u_1 = const\cdot (1-U)/(1+U)$. For $3 < q < \infty$, this
motivates the following test functions:
\begin{equation}
u_1 = \frac{1-U}{1 + \frac{q+1}{q-1} U}\; .
\label{Eq:TestF1}
\end{equation}
This yields
\begin{displaymath}
Z = \int_0^1 \frac{(1-U)^{-\frac{2}{q-1}}}{\left[ 1 + \frac{q+1}{q-1} U \right]^2}\, dU 
  \leq \int_0^1 (1-U)^{-\frac{2}{q-1}} dU = \frac{q-1}{q-3}
\end{displaymath}
and $E = -1/(2q)$. Therefore,
\begin{equation}
G_1 \leq (q-1)^2\frac{E}{Z} \leq -\frac{(q-1)(q-3)}{2q} < 0.
\end{equation}
Asymptotically as $q\rightarrow\infty$, we have $Z \simeq 1/2$, and
the test function $u_1$ defined in Eq. (\ref{Eq:TestF1}) becomes
optimal in this limit, so we expect that for $q >> 1$,
\begin{equation}
G_1 \simeq -q.
\end{equation}
The numerical results in the next subsection will confirm this behavior.

For $q=2$ or $q=3$, the decay of $u_1$ defined in
Eq. (\ref{Eq:TestF1}) as $U$ approaches $1$ is not rapid enough for
$Z$ to be finite. For this reason, we define $u_1 = (1-U)^m/[ 1 +
(q+1)U/(q-1)]$ with $m=3$ for $q=2$ and $m=2$ for $q=3$. These test
functions yield the estimates
\begin{eqnarray}
G_1 &\leq& -\frac{1}{9} \frac{381 - 512\log(2)}{15 - 16\log(2)} \approx -0.74
\qquad\hbox{for $q=2$,}\\
G_1 &\leq& -2\frac{4 - 3\log(3)}{2 - \log(3)} \approx -1.56
\qquad\hbox{for $q=3$.}
\end{eqnarray}
As we will see in the next subsection, these bounds are very close to
the actual eigenvalue.

To summarize, we have shown that the pulsation operator defines a
self-adjoint operator whose spectrum is contained in the interval
$[-2(q^2-1),\infty)$. We have also shown that the lower limit of the
spectrum is negative, and is independent of $\epsilon$. In particular,
this proves that for all $0\leq \epsilon < 1$, $q\geq 2$, the static
bubble solutions are unstable. In the next subsection we show the
existence of a unique non-degenerated negative eigenvalue for each $q
\geq 2$.

\subsection{Quantitative analysis -- eigenvalues of ${\cal A}$}

We consider the eigenvalue problem
\begin{equation}
-\left( U\bar{R}^q w' \right)' - \frac{2(q^2-1)}{\bar{R}^q\left[ 1 + \frac{q+1}{q-1} U \right]^2}\, w = \lambda\bar{R}^q w,
\label{Eq:Sturm-Liouville}
\end{equation}
where here and in the following, a prime denotes differentiation with
respect to $\bar{R}$, and $w$ is a smooth function on
$(1,\infty)$ such that all its derivatives are bounded near
$\bar{R}=1$ and decay faster than $\bar{R}^{-q/2-1/2}$ near
$\bar{R}=\infty$. We show that there exists precisely one negative
eigenvalue $\lambda$ to this problem and compute it by numerical
means. The following analysis is standard and follows the theory of
singular Sturm-Liouville problems \cite{JoergensRellich}.

We first show that there exists local solutions with the required
behavior near $\bar{R}=1$ and $\bar{R}=\infty$.  At the location of
the bubble ($\bar{R}=1$), we have a regular singular point while at
$\bar{R}=\infty$ we have an essential singularity. In order to analyze
the behavior near $\bar{R}=1$, it is convenient to replace the
coordinate $\bar{R}$ by $U = 1 - \bar{R}^{-(q-1)}$. We then obtain the
system
\begin{equation}
U\frac{\partial\Phi}{\partial U} = A(U) \Phi,
\qquad \Phi = \left( \begin{array}{c} w \\ U\frac{\partial w}{\partial U} \end{array} \right),
\label{Eq:AU}
\end{equation}
where
\begin{displaymath}
A(U) = \left( \begin{array}{cc} 0 & 1 \\
 - 2\frac{q+1}{q-1}\frac{U}{\left[1 + \frac{q+1}{q-1} U \right]^2}- \frac{\lambda U}{(q-1)^2} (1-U)^{-\frac{2q}{q-1}} & 0
\end{array} \right).
\end{displaymath}
For $0 \leq U < 1$, $A(U) = \sum_{k\geq 0} A_k U^k$ is analytic in $U$
with
\begin{displaymath}
A_0 = \left( \begin{array}{cc} 0 & 1 \\ 0 & 0 \end{array} \right),
\qquad
A_1 = \left( \begin{array}{cc} 0 & 0 \\ -2\frac{q+1}{q-1} - \frac{\lambda}{(q-1)^2} & 0 \end{array} \right),
\qquad
A_2 = ...
\end{displaymath}
Therefore, to leading order in $U$, Eq. (\ref{Eq:AU}) reads
\begin{displaymath}
\frac{\partial\Phi}{\partial(\log U)} = A_0\Phi,
\end{displaymath}
with solution $\Phi_0 = (1 + A_0\log U) e_0$, where $e_0$ is a
constant vector. Since we require $\Phi$ to be bounded near $U=0$ we
have to choose $e_0 = (1,0)$ in order to avoid the logarithmic term.
According to the general theory of regular singular points
\cite{Walter}, this gives rise to a one-parameter family of local
solutions near $U=0$ of the form $\Phi = \sum_{k\geq 0} e_k
U^k$. Plugging this form into Eq. (\ref{Eq:AU}) yields the recursion
relation
\begin{equation}
e_k = (k{\bf I} - A_0)^{-1} \sum\limits_{l=1}^k A_l e_{k-l}\; , \quad k = 1,2,3,...\; ,
\end{equation}
where ${\bf I}$ denotes the identity matrix. The series converges for
$U > 0$ small enough \cite{Walter}.

In order to analyze the behavior near $\bar{R}=\infty$, we first introduce
the coordinate $\rho$ defined by
\begin{equation}
\rho(\bar{R}) = \int_{1}^{\bar{R}} \frac{dR}{\sqrt{U(R)}}\; ,\quad
\bar{R} \geq 1
\end{equation}
which runs from zero to infinity. Notice that $\rho/\bar{R}
\rightarrow 1$ as $\bar{R} \rightarrow\infty$. Therefore, there is a
smooth function $h(\rho)$ such that $h(\rho)\rightarrow 1$ as $\rho
\rightarrow\infty$ and such that $\rho = h(\rho)\bar{R}$. In terms of
this new coordinate, we can rewrite the eigenvalue problem
(\ref{Eq:Sturm-Liouville}) as
\begin{equation}
\frac{\partial^2}{\partial\rho^2} X + \left( \lambda + B(\rho) \right) X = 0,
\qquad
B(\rho) = -\frac{1}{T}\frac{\partial^2 T}{\partial\rho^2}
          + \frac{2(q^2-1)}{\bar{R}^{2q}\left[ 1 + \frac{q+1}{q-1} U \right]^2}
\label{Eq:SLInfty}
\end{equation}
where $X = T w$, $T = \bar{R}^{q/2} U^{1/4}$. Since $B(\rho) =
O(\rho^{-2})$ it follows from a theorem by Dunkel \cite{Dunkel} that
there exists local solutions near $\rho = \infty$ of the form
\begin{equation}
X(\rho) = e^{-\Omega\rho} \left[ 1 + c(\rho) \right],
\label{Eq:X}
\end{equation}
where $\Omega = \sqrt{-\lambda}$ and where $c(\rho)$ is a twice
continuously differentiable function which satisfies $c(\rho)
\rightarrow 0$ and $\partial_\rho c(\rho) \rightarrow 0$ as
$\rho\rightarrow\infty$. The introduction of the ``tortoise''
coordinate $\rho$ is important for the case $q=2$, where
\begin{displaymath}
\rho = \sqrt{\bar{R}}\sqrt{\bar{R}-1} + \log\left(  \sqrt{\bar{R}} + \sqrt{\bar{R}-1} \right),
\end{displaymath}
and so $e^{-\Omega\rho} = \bar{R}^{-\Omega/2} e^{-\Omega\bar{R}}
\left[ b + \tilde{c}(\bar{R}) \right]$, where $b = 2^{-\Omega}
e^{\Omega/2}$ and $\tilde{c}(\bar{R})$ is twice continuously
differentiable and satisfies $\tilde{c}(\bar{R}) \rightarrow 0$,
$\partial_{\bar R}\tilde{c}(\bar{R}) \rightarrow 0$ as
$\bar{R}\rightarrow\infty$. For $q \geq 3$ one can show that $h = 1 +
O(\bar{R}^{-1})$, and in those cases, one can simply replace
$e^{-\Omega\rho}$ by $e^{-\Omega\bar{R}}$ in the expression
(\ref{Eq:X}).

Therefore, there exists local solutions $w_L,w_R$, near left and right
 boundary points respectively, of the form
\begin{eqnarray}
w_L(\lambda; \bar{R}) &=& 1 - \left( 2\frac{q+1}{q-1} + \frac{\lambda}{(q-1)^2}\right) U + O(U^2),
\label{Eq:LeftBC}\\
w_R(\lambda; \bar{R}) &=& U^{-1/4} \bar{R}^{-q/2} e^{-\sqrt{-\lambda}\rho} \left[ 1 + c(\lambda; \rho) \right],
\label{Eq:RightBC}
\end{eqnarray}
where $c(\lambda; \rho)$ is a twice continuously differentiable
function which satisfies $c(\lambda; \rho) \rightarrow 0$ and
$\partial_\rho c(\lambda; \rho) \rightarrow 0$ as
$\rho\rightarrow\infty$. By varying $\lambda$ in the range of negative
real values, we can look for a global solution by the ``shooting and
matching'' procedure.

In what follows, we will prove that there exists such a global
solution and that it is unique up to a multiplicative factor. The idea
is to use the fact that the local solution at $\bar{R}=1$ for
$\lambda=0$, which is given explicitly by
\begin{equation}
w_L(0; \bar{R}) = w_0(\bar{R}) \equiv \frac{q-1 - (q+1)U}{q-1 + (q+1)U}\; ,
\end{equation}
has exactly one node in the interval $0 < \bar{R} < \infty$ and use
the nodal theorem which states that the number of bound states is
equal to the number of nodes of the local solution belonging to zero
eigenvalue. For a precise statement of this theorem, which also
applies to coupled systems of Schr\"odinger equations, we refer to
\cite{AQ-Nodal}.  Below, we provide an elementary proof for the
existence of a unique negative eigenvalue. This proof consists of
three parts: (i) To show that the eigenvalues are non-degenerate
assuming they exist; (ii) that an eigenvalue exists, by demonstrating
that the `shooting and matching' procedure has a solution and (iii)
that the eigenvalue is unique.\\

In order to proof these assertions we rely on the following
observations.  Let $w_1$ and $w_2$ be two local solutions to the
eigenvalue problem (\ref{Eq:Sturm-Liouville}), belonging to the
eigenvalues $\lambda_1$ and $\lambda_2$, respectively. The Wronski
determinant is defined by
\begin{equation}
W[w_1,w_2] = U \bar{R}^q\left( w_1' w_2 - w_2' w_1 \right) \; ,
\label{Eq:Wronski}
\end{equation}
and by virtue of Eq. (\ref{Eq:Sturm-Liouville}), it satisfies
\begin{equation}
W'[w_1,w_2] = (\lambda_2 - \lambda_1) \bar{R}^q w_1 w_2\; .
\label{Eq:WronskiProp}
\end{equation}
The Wronski determinant will play a central role in proving the three
statements above.\\

(i) {\em The eigenvalues are non-degenerate}.  Suppose there are two
eigenfunctions, $w_1$, $w_2$ belonging to the same eigenvalue
$\lambda_1 = \lambda_2$. Then, Eq. (\ref{Eq:WronskiProp}) implies that
$W[w_1,w_2]=const$. But this constant is zero since $w_1$, $w_2$ and
their derivatives are bounded near $\bar{R}=1$, and because of the
factor $U$ in the definition of the Wronski determinant. Therefore,
$W[w_1,w_2] \equiv 0$, which means that $w_1$ and $w_2$ are linearly
dependent. \\

(ii) {\em There exists a negative eigenvalue}.  Let $\bar{R}_1 > 1$ be
large enough such that for each $-2(q^2-1)-1 \leq \lambda \leq 0$ and
all $\bar{R} \geq \bar{R}_1$, $w_R(\lambda; \bar{R}) > 0$,
$w_R'(\lambda; \bar{R}) < 0$ and $w_0(\bar{R}) < 0$. That we can
always choose such a $\bar{R}_1$ follows from the definition of $w_0$
and the behavior of $w_R$ near the right boundary. Additionally, we
can extend $w_L(\lambda; \bar{R})$ to $\bar{R}=\bar{R}_1$. Now define
$W(\lambda) = \left. W[w_L(\lambda),w_R(\lambda)]
\right|_{\bar{R}=\bar{R}_1}$. We can match $w_L(\lambda)$ and
$w_R(\lambda)$ if and only if $W(\lambda)=0$. So our task it to show
that the function $\lambda \mapsto W(\lambda)$, which is continuous,
possesses a zero. Now choose $\lambda = \lambda_1 \equiv
-2(q^2-1)-1$. Since
\begin{equation}
w_L(\lambda_1; \bar{R}) = 1 + \frac{U}{(q-1)^2} + O(U^2),
\end{equation}
we have $w_L(\lambda_1; 1) = 1$, $w_L'(\lambda_1; 1) >
0$. Furthermore, it follows from Eq. (\ref{Eq:Sturm-Liouville}) that
for $\lambda < -2(q^2-1)$, $w(\bar{R})$ cannot have a local maxima as
long as $w > 0$. Therefore, $w_L(\lambda_1; \bar{R})$ is monotonously
increasing and we must have $w_L(\lambda_1; \bar{R}_1) > 0$,
$w_L'(\lambda_1; \bar{R}_1) > 0$.  This implies that $W(\lambda_1) >
0$. On the other hand, for $\lambda = 0$, we have $w_L(0;\bar{R}_1) <
0$, $w_L'(0;\bar{R}_1) < 0$ which implies $W(\lambda_1) <
0$. Therefore, there exists $\lambda$ with $\lambda_1 < \lambda < 0$
for which $W(\lambda)=0$ and for which the local solutions
$w_L(\lambda)$ and $w_R(\lambda)$ can be matched to yield an
eigenfunction $w_\lambda(\bar{R})$ belonging to the negative
eigenvalue $\lambda$.\\

(iii) {\em The negative eigenvalue is unique}. In order to prove this,
we first show that $w_\lambda(\bar{R})$ possesses no nodes in
$(1,\infty)$. We normalize $w_\lambda$ such that
$w_\lambda(1)=1$. Suppose that $w_\lambda$ has a node at $\bar{R}_2 >
1$, and that $w_\lambda$ is strictly positive on
$(1,\bar{R}_2)$. Then, $w_\lambda'(\bar{R}_2) \leq 0$. Integrating
Eq. (\ref{Eq:WronskiProp}) from $\bar{R}=1$ to $\bar{R}=\bar{R}_2$ we
have, choosing $\lambda_1=\lambda$ and $\lambda_2=0$,
\begin{equation}
\left[ U\bar{R}^q w_\lambda' w_0 \right]_{\bar{R}=\bar{R}_2}
 = W[w_\lambda,w_0](\bar{R}_2) - W[w_\lambda,w_0](1) 
 = -\lambda\int_1^{\bar{R}_2} \bar{R}^q w_\lambda w_0 d\bar{R}.
\end{equation}
It follows that $w_0(\bar{R}_2) < 0$ for otherwise, the left-hand side
would be negative or zero and the right-hand side positive. Therefore,
$\bar{R}_2$ lies to the right of the single node of $w_0$. Next,
integrate Eq. (\ref{Eq:WronskiProp}) from $\bar{R}=\bar{R}_2$ to
some $\bar{R} > \bar{R}_2$ and obtain
\begin{equation}
U\bar{R}^q w_\lambda'(\bar{R}) w_0(\bar{R}) = -\lambda\int_{\bar{R}_2}^{\bar{R}} \bar{R}^q w_\lambda w_0 d\bar{R}
 + \left[ U\bar{R}^q w_\lambda' w_0 \right]_{\bar{R}=\bar{R}_2} 
 + U\bar{R}^q w_\lambda(\bar{R}) w_0'(\bar{R}).
\end{equation}
Suppose first that $w_\lambda < 0$ on $(\bar{R_2},\bar{R})$. Then, the
right-hand side is positive, so we must have $w_\lambda'(\bar{R}) <
0$. On the other hand, if $w_\lambda > 0$ on $(\bar{R_2},\bar{R})$ the
right-hand side is negative, and we must have $w_\lambda'(\bar{R}) >
0$.  Both options contradict the fact that $w_\lambda(\bar{R})
\rightarrow 0$ as $\bar{R} \rightarrow \infty$. Therefore, $w_\lambda$
is positive on $(1,\infty)$ and possesses no nodes. Finally, suppose
now there exists another eigenfunction $w_{\hat{\lambda}}$ belonging to a
negative eigenvalue $\hat{\lambda} > \lambda$. Integrating
Eq. (\ref{Eq:WronskiProp}) over $(1,\infty)$ and taking into account
the boundary conditions, it follows choosing $\lambda_1=\lambda$ and
$\lambda_2=\hat{\lambda}$ in Eq. (\ref{Eq:WronskiProp}), that
\begin{equation}
0 = \left. W[w_\lambda,w_{\hat{\lambda}}] \right|_1^\infty
  = (\hat{\lambda} - \lambda)\int_1^\infty \bar{R}^q w_\lambda w_{\hat{\lambda}} d\bar{R},
\end{equation}
which implies that $w_{\hat{\lambda}}$ must have at least one node. But
then we can repeat the argument in the first part of (iii) and reach a
contradiction by replacing $w_\lambda$ by $w_{\hat{\lambda}}$.

Therefore, there exists a unique non-degenerated negative
eigenvalue. Although we have not proven that this eigenvalue
corresponds to the ground state, we conjecture that the spectrum of
the pulsation operator consists of the single negative eigenvalue and
a continuous spectrum $[0,\infty)$.

We have computed the eigenvalue numerically following the ``shooting
and matching'' procedure described in \cite{NumRec}. In order to do
so, we rewrote equation (\ref{Eq:SLInfty}) in terms of the variable $Y
= e^{\Omega\rho} X$ and implemented it numerically. To proceed with
the numerical integration we introduce the compactified coordinate
$U=1-\bar R^{-(q-1)}$ and define left/right boundaries at $U=0+\Delta$
and $U=1-\Delta$ respectively.  At these boundaries, initial values
are given in correspondence with the local solutions
(\ref{Eq:LeftBC}), (\ref{Eq:RightBC}), respectively, where the left
local solution is truncated after the second order term in $U$ and
where $c(\rho)=0$ is chosen in the right solution. The equations are
then integrated using the LSODE package \cite{lsode} combined with a
nonlinear Newton iteration to match the solutions at an intermediate
point by modifying the parameter $\lambda$ and the scaling of the
right solution. The tolerance specified in the LSODE solver, which is
used in the variable time-step ODE integrator, was set to
$\mbox{TOL}=10^{-13}$. Additionally, we require that at the
intermediate point, {\it $Y$ and its derivative} match with an error
smaller than $10^{-9}$. \\

\begin{figure}
\begin{center}
\epsfig{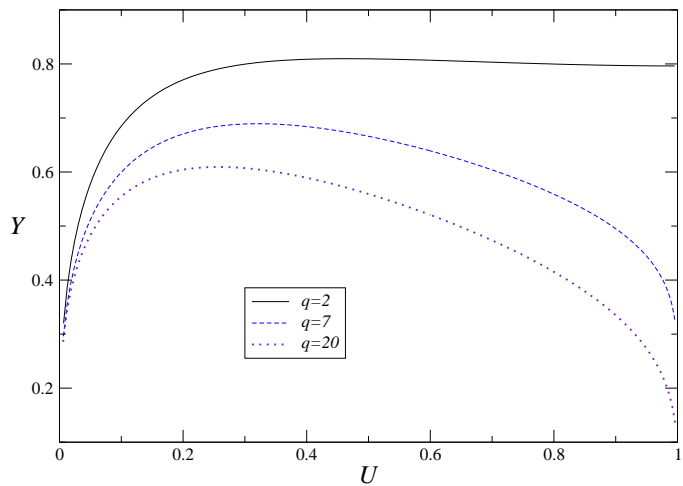}
\end{center}
\caption{The eigenfunction $Y = e^{\Omega\rho} X$ as a function of $U$
for three different values of $q$.}
\label{Fig:func_q}
\end{figure}

As an illustration, figure \ref{Fig:func_q} presents the obtained
solutions $Y = e^{\Omega\rho} X$ for the cases $q=2,7,20$.  The
matching point was chosen at $U=0.6$ but we checked that the obtained
results are independent of this choice.\\

We then use the developed code to obtain the dependence of $\Omega$ on
the dimensionality $q$ in the range $q = 2,3,...250$. To ensure
convergence of the obtained value we run the code with $\Delta =
0.01/2^n$ ($n$ an integer number) and increase $n$ until the
difference of two successive values of $\Omega$ lies below
$0.01$\%. The results for $\Omega$ are presented in figure
\ref{Fig:lam_q}, while figure \ref{Fig:error_q} displays the
percentile error $PE=100\, |\sqrt{q}-\Omega|/\sqrt{q}$. As expected,
the discrepancy diminishes as $q$ increases.

\section{Final Words}
\label{Sect:Conclusions}

This work has put on firm grounds the existence of critical phenomena
in bubble spacetimes confirming the indications obtained via numerical
simulations. We find spherically symmetric, in $z$ homogeneous, linear
modes that depart like $\exp(\Omega t/R_0)$ from the critical solution
which is static\footnote{Recently, an independent analysis, also styding gauge-independent 
linearized perturbations off the critical solution for the case $q=2$ 
has obtained values in full agreement with the ones presented here\cite{hovdebo}).}.
We have identified the critical solution, the growth
rate of the unstable mode and established the universality of the
growth rate. The universality follows from the fact that it is
possible to decouple the pulsation equations into two Sturm-Liouville
problems, where one problem does not give rise to instabilities while
the other one is manifestly independent of the background solution
when expressed in terms of appropriate dimensionless
quantities. Additionally, the study of the unstable Sturm-Liouville
problem allows for a derivation of the simple relation $\Omega \simeq
q^{1/2}$ in the limit of large $q$'s. This relation agrees very well
with the numerical results. 

Furthermore, we have made use of a double analytical continuation to
map the solution to a stationary mode of the corresponding black
strings which is proportional to $e^{i\Omega z/R_0}$. This mode
represents a static deformation of the black string, with harmonic
dependency in $z$. This mode is allowed if the asymptotic periodicity
of the extra dimension is an integer multiple of $L_c = 2\pi
R_0/\Omega$. For the uncharged case, the obtained values of $L_c$ give
the the critical length for the Gregory-Laflamme instability\cite{GL,
GL2}.  In the charged case, the black string solutions given by
(\ref{Eq:BubbleSol1},\ref{Eq:BubbleSol2}) do not correspond to those
considered in \cite{GL2} since we restricted ourselves to a ${\bf U(1)}$ gauge field
in the absence of a dilaton.  We give the
dependency for the critical length in terms of the parameter
$\epsilon$ in an analytic way. This result can be expressed in terms
of the dimensionless mass introduced in\cite{sorkin}
\begin{equation}
\mu := \frac{M_c}{L_c^q} = \frac{|S^q|}{16\pi} \left( \frac{\Omega}{2\pi} \right)^{q-1}
\left[ q + 2\epsilon\left(q - \frac{1}{q} \right) \right].
\end{equation}
For large $q$ we can approximate $\Omega \simeq q^{1/2}$ and find
\begin{equation}
\mu \simeq \frac{\sqrt{q}}{16} \left( \frac{e}{2\pi} \right)^{q/2} (1 + 2\epsilon),
\end{equation}
which provides a different way of deriving the law presented in
Ref. \cite{sorkin,sorkinkol}, and generalizes it to the charged case.

Note there is a remarkable difference between the instability of the
charged strings considered here and those in \cite{GL2,GL3}. Whereas
the addition of charge makes the critical length, at which the instability
arises, shorter in our case --making the string ``more unstable''-- 
while longer (and the extremal case marginally stable) in \cite{GL2,GL3}. 
The reason for such a marked different is likely due to the interaction with
a magnetic field in \cite{GL2,GL3} while an electric field in the present case.
Thus, the dynamics of charged strings even in 5-dimensional cases appears
to be quite rich.

Finally, (single) Wick rotations are often used to study and understand phenomena
in the Euclidean sector and draw conclusions about black branes. Here,
and also in \cite{elvangharmarkobers}, it is show
 that the understanding of 
solutions of bubble spacetimes can be used to shed light into black string
systems and viceversa.

\section{Acknowledgments}
We wish to give special thanks to H. Beyer for many useful discussions
and for pointing out to us Ref. \cite{Dunkel}, R. Myers for
continuous discussions on bubbles and strings and R. Emparan for 
several valuable comments and suggestions. We also acknowledge
interesting discussions and comments with M. Choptuik, J. Hovdebo, 
D. Neilsen, F. Pretorius,
J. Pullin and J. Ventrella.  This work was supported in part by the Center for
Computation \& Technology at Louisiana
 State University, by 
 grants NSF-PHY-0244335, NSF-PHY-0244299, NSF-INT-0204937, NASA-NAG5-13430 and a Research
Innovation Award from Research Corporation to Louisiana State University and by funds
from the Horace Hearne Jr. Laboratory for Theoretical Physics. We
thank the National University of Cordoba and L.L. thanks the Perimeter
Institute for Theoretical Physics for hospitality where parts of this
work were completed. L.L. is an Alfred P. Sloan Fellow.

\end{document}